\documentclass[copyright,creativecommons]{eptcs}
\providecommand{\event}{DCM 2010} % Name of the event you are submitting to
\usepackage{breakurl}             % Not needed if you use pdflatex only.
\usepackage[dvips]{graphics,color}
\usepackage{graphicx}
\usepackage{amsmath}
\usepackage{amsfonts}

\usepackage{amssymb}
\usepackage{amstext}

\title{Measurement Based Quantum Computation on Fractal Lattices}
\author{Damian Markham
\institute{CNRS, LTCI, Telecom ParisTech, 37/39 rue Dareau, 75014 Paris, France}
\and
Janet Anders
\institute{Department of Physics \& Astronomy, University College London, London WC1E 6BT, UK}
\and
Michal Hajdu\v{s}ek
\institute{The School of Physics and Astronomy, University of Leeds, Leeds, LS2 9JT, United Kingdom}
\and
Vlatko Vedral
\institute{Centre for Quantum Technologies, National University of Singapore, Singapore}
\institute{Department of Physics, National University of Singapore, Singapore}
\institute{Clarendon Laboratory, University of Oxford, Oxford UK}
}
\def\titlerunning{Measurement Based Quantum Computation on Fractal Lattices }
\def\authorrunning{D.~Markham, J.~Anders, M.~Hajdu\v{s}ek and V.~Vedral}
\begin{document}
\maketitle

\begin{abstract}
In this article we extend on work which establishes an analogy between one-way quantum computation and thermodynamics to see how the former can be performed on fractal lattices. We find fractals lattices of arbitrary dimension greater than one which do all act as good resources for one-way quantum computation, and sets of fractal lattices with dimension greater than one all of which do not. The difference is put down to other topological factors such as ramification and connectivity. This work adds confidence to the analogy and highlights new features to what we require for universal resources for one-way quantum computation.\end{abstract}

\section{Introduction}

Drawing analogies is often a very powerful tool in science. It can allow not only deepened understanding through new perspectives opened up, but it can also allow technical tools from one discipline to be applied to another, often very fruitfully. In \cite{Anders07} an analogy between measurement based quantum computation and thermodynamics is made, viewing the computation itself as a phase transition. This is in spirit reverse to the thought provoking analogy made by Toffoli \cite{Toffoli98} where physics is viewed as a computation, rather \cite{Anders07} tries to understand computation itself as a physical process. In doing so key features of useful resources for one-way quantum computers were identified, in direct analogy to the identification of critical systems in thermodynamics. In particular rather elegant and simple methods first developed by Peierls \cite{Peierls36} to show that one dimensional spin chains are not critical, where as two dimensional chains are, were translated into arguments of the dependence of dimension on universal resources for one-way quantum computation. In this work we extend this theme to look for other important features of universal resources, following the work on fractal lattices by Gefen et. al. \cite{Gefen80}. There it is shown that critical behaviour in spin systems relies not just on the dimension of the lattice, but also other features such as order of ramification and connectivity. We again see an exact mirroring of results, highlighting these also as features crucial for universal resources for one-way quantum computation. As examples we will see that there exists a set of fractal lattices (Sierpinski carpets) for which any dimension greater than one guarantees it can act as a universal resource. On the other hand we will also see examples of dimension greater than one which are not universal, highlighting the importance of the other topological features (ramification, connectivity and lacunarity).

\section{The analogy: Phase Transition and Measurement Based Quantum Computation}

We start by reviewing the problems addressed in this analogy. In the case of thermodynamics and many-body physics, the problem which is of interest is the existence, or not, of some critical phenomena or phase transition. Simply put, a phase transition is when a small change in some parameters of a given system gives rise to a large macroscopic change of state, or phase. For example, at just below zero degrees water becomes ice, and just above it becomes water again. These two phases of matter are clearly very different. In spin systems the macroscopic property of interest is whether the system is magnetised or not. This happens when sufficiently many spins point in the same direction - we call this the `ordered state'. The effect is witnessed by the amount of magnetisation $M$ present - which is called an `order parameter'. In the Ising model the ground state (corresponding to zero temperature) is ordered, and for high temperatures, the orientation of the spin becomes random and it is not ordered - its magnetisation is zero. The question is then whether or not there is a finite, non-zero, temperature $T_{crit}$ below which the system is ordered. If this is the case we say there can be a phase transition from non-magnetised to magnetised at temperature $T_{crit}$. It is known that for one-dimensional spin chains with nearest neighbour interactions only, there is no phase transitions, where as for two dimensional lattices there are. This will be explained via the Peierls argument below.

In the case of measurement based quantum computation, the problem of interest is the ability, or not, to perform universal quantum computation. In one-way quantum computation (1-way QC, in this work we take it to be synonymous with measurement based quantum computation) \cite{Raussendorf01}, a computation is carried out, first by preparing a highly entangled multipartite quantum state (which we call a `resource state', and which is independant of the actual computation to be performed) and then performing local measurements and local corrections on individual sites. The choice of measurements, and how they depend on each other determines the computation which is performed. During the process of measurement entanglement is destroyed, and in this sense consumed by the computation. At the end of the computation, the classical information $I$ of the solution is obtained as the measurement outcomes of the last few measurements. Since its invention a large amount of effort has gone into finding out what constitutes a good initial resource state (see e.g. \cite{VandenNest07,Gross09}). Given a particular set of states, the question is, whether or not it can act as a universal resource for quantum computation. The analogy we will now draw goes towards answering this question. Note that we will always consider resource states as graph states in this work.

\begin{figure}[h]
    \begin{center}
    \includegraphics[width=0.35\textwidth]{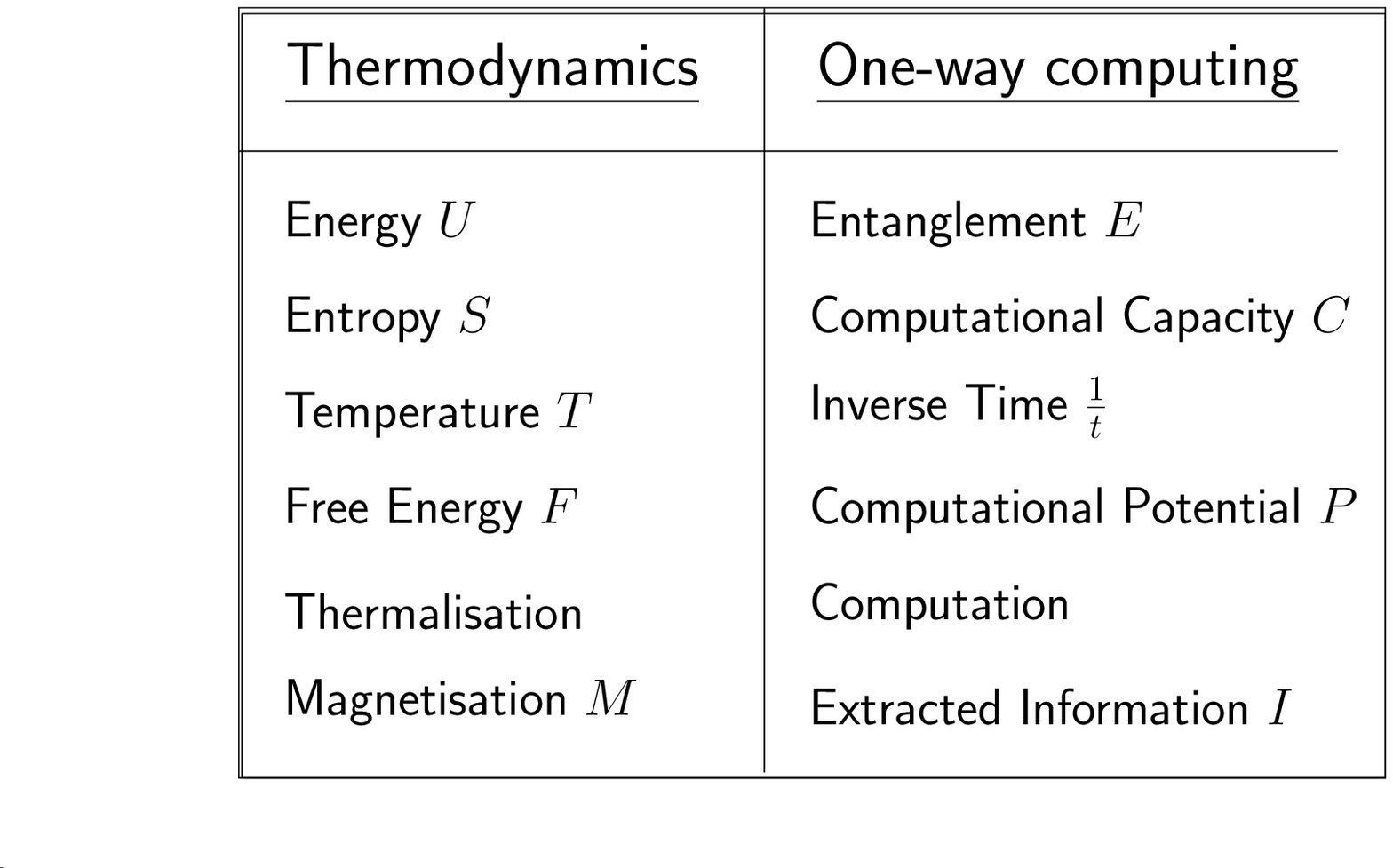}
    \caption{\label{fig:comparison} (Borrowed from \cite{Anders07}) On the left, a set of quantities from thermodynamics and on the right their analogous counterparts for one-way quantum computation.}
    \end{center}
\end{figure}

A list of the analogous quantities between thermodynamics on the one hand, and one-way quantum computation on the other is given in Fig.~\ref{fig:comparison}.

The second law of thermodynamics states that systems interacting with a thermal bath will always conspire to minimise the free energy $F$, given by
\begin{align} \label{eqn: 2nd Law}
F=U-TS.
\end{align}
Intuitively we can understand the second law as saying that by the process of thermalisation, nature insists that at a given temperature $T$, the energy $U$ is spread out as much as possible, by maximising the entropy $S$.

When making our analogy the quality we are insisting upon for our 1-way QC computation is that it should be universal. To this end,  we postulate a kind of `law of 1-way QC' whereby we insist (as `mother nature' of the 1-way QC - it is after all us who designs and controls it) that the quantum computer be as universal at each step as possible. That is, we insist that at each time the computation is carried out in such a way as to maximise the potential. In terms of quantities, we say that for a given amount of entanglement $E$, we insist that any computation at any time $t$ maximises the number of ways it can be used - which we call the computational capacity $C$. We thus phrase our `second law of universal 1-way QC' as that at each time $t$, the potential $P$, given by
\begin{align}  \label{eqn: 2nd Law of QC}
P=E-1/tC
\end{align}
must be minimised (i.e. the potential should be consumed as fully as possible).

Following an argument first put forward by Peierls \cite{Peierls36}, and cleaned up by Griffiths \cite{Griffiths64}, which shows that one dimensional spin chains are not critical, but two dimensional spin chains are, an intuitive argument as to why a one dimensional cluster state is not a universal resource for 1-way QC, where as a two dimensional cluster state is a universal resource was given \cite{Anders07}.
Peierls' argument goes as follows. If we want to test whether an `ordered state' (i.e. one with a large number of spins pointing in the same direction such that there is overall a positive magnetisation) is possible for some nonzero temperature $T$, we simply check wether small pertubations to this state will raise or lower the free energy of that state (the very physicsy `shake it and see' approach). Any such pertubation will change the free energy
\begin{eqnarray} \label{eqn: Pertbn of F}
\Delta F = \Delta U - T \Delta S,
\end{eqnarray}
If by perturbing it we can {\it reduce} the free energy, the state clearly is not a valid thermal state, by the 2nd law. In terms of equation (\ref{eqn: Pertbn of F}) this is then a question of balance between the change in energy and the change in entropy. If perturbing the system increases the entropy more than the energy, the state before pertubation was not a valid thermal state. In the case of a one dimensional spin chain, the cost of any pertubation in terms of entropy is much greater (it scales with the number of spins $n$) than the cost in energy (which is fixed). In the case of a two dimensional spin, they scale with $n$ in the same way, hence a balance can be found. By finding the fixed point of the free energy (the point where the pertubation makes no change - by setting (\ref{eqn: Pertbn of F}) to zero), a critical temperature can be found above which the system is not ordered. Remarkably, given the simplicity of this approach, this is very close to the actual critical temperature, below which it can be shown also that the system is ordered.

When testing if a system can be used for 1-way QC, the ordered state is the `solution state' (the state after all measurements have been made in the 1-way QC) and the test is, if it is possible for some finite time $t$. Again, we test this by perturbing it and seeing if it violates our 2nd law of 1-way QC. If it does, it is definitely not a valid state according to our second law - that is, no computation satisfying our `law of 1-way QC' can find such a state at a time $t$. Any partubation results in a change in computational potential
\begin{eqnarray} \label{eqn: Pertbn of P}
\Delta P = \Delta E - 1/t \Delta C.
\end{eqnarray}
This then bares out as a balance between the entanglement $E$ and the number of ways of using it $C$, for a given $t$. As above, in the case of a one dimensional cluster state, this is balance can not be met - the number of ways of using the entanglement is larger than the entanglement available, hence some choice must be made about its use, sacrificing universality. Alternatively it says that there is no finite time length at which it could be achieved, so if it were possible, it would take an infinite amount of time. On the other hand as for the spin case, a two dimensional lattice these quantities do balance. Again as above it is possible to approximate a critical time $t_{crit}$ below which the computation cannot be completed in a universal fashion, by setting by setting (\ref{eqn: Pertbn of P}) to zero. In fact by seeing how both of these factors scale with dimension $D$, it is possible to arrive at the following formula
\begin{eqnarray}
t_{crit} \propto \frac{ln{D}}{D},
\end{eqnarray}
which agrees with both our intuition and examples that higher dimensional states can allow for greater speed in computation.

\section{Computing on Fractal Lattices}

We can now extend this analogy to cover another interesting set of examples from many-body physics, where it is shown that not only does dimension play a role in spin criticality, but also other topological features. In \cite{Gefen80} similar techniques to those of Peierls and Griffiths described above are used to test the criticality of spin systems, this time based over several self-similar fractal lattices. Again the arguments are testing the ability of a lattice to balance the change in energy and entropy for small pertubations. Examples are presented which both do and do not allow criticality for all (fractal) dimensions greater than one. The additional features which capture the existence of criticality are shown to be topological including ramification, connectivity and lacunarity.

We follow the same analogy as before to show exact mirrors of these results in 1-way QC. We see that graph states of the fractal lattices of Koch curves and Sierpinski gaskets are not universal resources for 1-way QC, where as Sierpinski carpets are, independent of dimension. As in \cite{Gefen80} we can interpret this as the role of other topological factors including the ramification.

\begin{figure}[h]
{\resizebox{!}{2cm}{\includegraphics{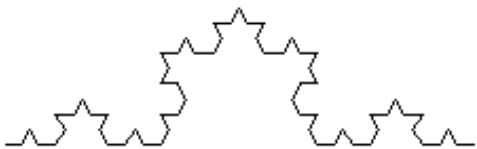}}} \caption{\label{FIG: KochCurve} The Koch curve. The dimension of the lattice is defined as $D=\frac{\ln{4}}{\ln{3}}=1.2619$. It has ramification $R=2$.}
\end{figure}

\begin{figure}[h]
{\resizebox{!}{4cm}{\includegraphics{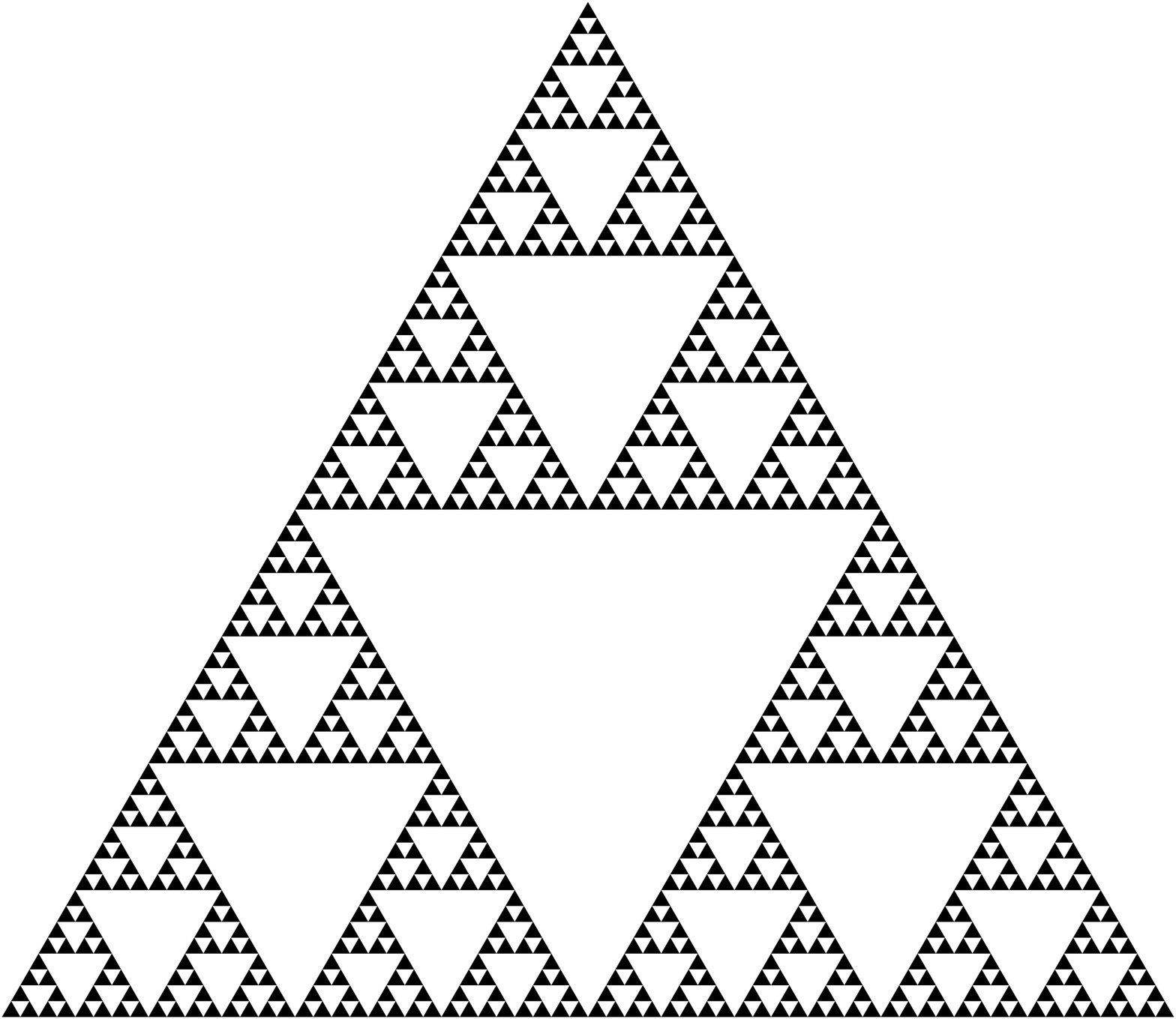}}} \caption{\label{FIG: SierpinskiGasket} A sierpinski gasket is constructed by subdividing an equilateral triangle into four subtriangles, then removing the central triangle and repeating the process. The dimension of the lattice is defined as $D=\frac{\ln{3}}{\ln{2}}=1.585$. It has ramification $R=3$ or $R=4$.}
\end{figure}

The Koch curve is illustrated in Fig.~\ref{FIG: KochCurve}. For our purposes this behaves exactly as in the 1D case in the previous section, where we argue it is indeed not a good universal resource \cite{Anders07}. Proofs to this effect are also known in the literature (\cite{VandenNest07}). The Sierpinski gasket is shown in Fig.~\ref{FIG: SierpinskiGasket}. Again, the same Peierls like arguments show it is not a valid possibility for a universal resource follow as those made above. That is the balance between the entanglement present and the number of ways to use the entanglement can not be found for some finite $t$. In analogy to the spin case \cite{Gefen80}, a significant pertubation of the solution state by adding entanglement can be done in many more ways than the amount of entanglement that is added, causing a negative change in P (equation (\ref{eqn: Pertbn of P})). This is unfortunately not a rigorous proof of non-universality, since our analogy (and in particular our `law of 1-way QC') is not proven, rather it is justified. We can however prove that the Sierpinski gasket is not a universal resource by methods introduced in \cite{VandenNest06}. There it is shown that if the entanglement does not scale with a family of resource states (such as our lattices), then it cannot be a universal resource for 1-way QC \cite{VandenNest06,VandenNest07}. The entanglement measure they use is the \textit{entanglement width} $E_{wd}$ defined as
\begin{eqnarray}
E_{wd}(|\psi\rangle):= \min_T \max_e E^{bi}_{T,e}(|\psi\rangle),
\end{eqnarray}
where $E^{bi}_{T,e}(|\psi\rangle)$ is the bipartite entropy of entanglement across the bipartite cut defined by $T$ and $e$. $T$ is a subcubic graph with $n$ leaves (edges not leading to a vertex at one end) and $e$ is an edge of $T$. Each leaf corresponds to a qubit. The bipartite cut is defined by removing edge $e$ to give two separate trees. The leaves of one tree correspond to one side of the cut, and the other tree the other side of the cut. It can easily be seen that for the Sierpinski gasket $|\psi_{SG}\rangle$ a tree can be defined with the same self similar properties, such that the best cut $e$ also has self similar properties and gives entanglement $E^{bi}_{T,e}(|\psi_{SG}\rangle)=3$ which does not grow. Hence the entanglement width is bounded $E_{wd}(|\psi_{SG}\rangle)\leq3$ for any lattice size. Since it does not scale with the size of the lattice, it cannot be a universal resource.

\begin{figure}[h]
{\resizebox{!}{4cm}{\includegraphics{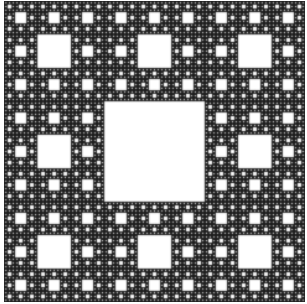}}} \caption{\label{FIG: Sierpinski} A Sierpinski carpet is constructed by subdividing a square into $b^2$ subsquares, then cutting out $l^2$ of these subsquares, and repeating the process. The dimension of the lattice is defined as $D=\frac{\ln{(b^2-l^2)}}{\ln{b}}$. Here we illustrate a Sierpinski carpet with $b=3$, $l=1$. The ramification is infinite.}
\end{figure}

On the other hand Sierpinski carpets (see Fig.~\ref{FIG: Sierpinski}) are universal for all dimensions greater than one. The arguments to show it is a valid possibility for a universal resource follow along the same lines as the Peierls like argument made above. That is the balance of entanglement present and the number of ways to use the entanglement can always be found for some finite $t$. This is of course not a proof that it is a universal resource, since, even if we assume our `law of 1-way QC', it only shows that it is a possible resource, i.e. that it doesn't violate the law of 1-way QC. We can however construct exact proofs for all cases. To show explicitly that these are universal, we adopt a similar technique to that used in \cite{Browne08}, which is to actively construct a standard 2D lattice by taking out vertices using local $Z$ and local $X$ measurements, which in turn is known to be a universal resource \cite{Raussendorf01}.
The idea is that, given an arbitrary lattice (which may even be irregular, as in the case of \cite{Browne08}), if we can draw a standard 2D grid over this lattice, we can measure away the extra qubits to leave only the ideal 2D lattice.
This is possible because of the way $X$ and $Z$ measurements convert one graph state to another (see e.g. \cite{Hein06}).
It is easy to see by looking at the Sierpinksi carpet Fig.~\ref{FIG: Sierpinski}, it is always possible to draw a 2D grid which grows with the size of the carpet. Thus we always have a way to get a known universal resource for any dimension greater than one.

We thus see that the ability of a lattice to act as a universal resource for 1-way QC does not just depend on dimension. In particular, for any dimension between $1$ and $2$ we can find a Sierpinski carpet which is a universal resource, where as we have seen two examples with dimension in this range which are not universal. As in \cite{Gefen80} we can then infer that it is down to other topological properties. One such property that resonates in the case of 1-way QC is \textit{ramification}. The ramification $R$ is the minimal number of edges that must be removed to separate a part of the lattice of arbitrary size. It tells us something about how globally connected the lattice is or how easily the lattice can be separated into chunks. The lower the ramification the easier it is to separate parts of the lattice off and the less globally connected it is. From our examples, those lattices with finite ramification are not universal resources, where as those with infinite ramification are (the 2D lattice also has infinite ramification). This is very similar in flavour to the idea behind the entanglement width introduced in \cite{VandenNest06} to check for universality of resource states (and used above). In a sense this also looks for some global connectedness, by the nature of the min max definition above. Here too an infinite scaling is required for universality. We may imagine there could be a connection between the two. We may also wonder whether an alternative entanglement measure can be defined with respect to ramification (or indeed other topological properties), which could be used to see their usefulness as resources for 1-way QC. This is beyond the scope of the current manuscript, but poses interesting possibilities.

\section{Conclusions}
We have seen that the analogy developed in \cite{Anders07} can be used to argue that fractal lattices can also act as universal resources for 1-way QC, and that not only is dimensionality important, but also other topological features such as ramification. By providing further examples where the analogy succeeds we have strengthened its validity. It also highlights new features that we can expect good resources should posses for 1-way QC. We can also ask how this corresponds to known conditions for good resources, such as the entanglement conditions in \cite{VandenNest07}. In this context perhaps it is possible that the important  topological features could also correspond to particular entanglement features. Another possible connection to existing conditions would be to the existence of Flow on these lattices. Flow (and gFlow) are known sufficient conditions for a lattice, or graph to allow 1-way QC \cite{Danos06,Browne07}. The fact that for example in the Sierpinski carpets we can always reduce them to a 2D lattice implies that we can always extract a flow in some sense. Perhaps the topological features presented here are also important for the existence of flow. We also note that the techniques, and indeed the lattices used are very similar to those in \cite{Browne08}, which arise in the context of 2D lattices with noise. In a sense this is no surprise since it is similar situations which may give rise to fractal lattices in many-body physics also. But it may also indicate that the analogy used could be useful in treating noise over fixed lattices. On foundational level, this analogy, and its reenforcement by this work, opens up many interesting questions and possibilities. For example, can these analogies be made more solid by a kind of path integral approach to 1-way QC? How deep can we take these analogies beyond 1-way QC, can it work for example for other models of computation? We hope this work will stimulate further research in these areas.

\bigskip
\noindent \textit{Acknowledgements} DM acknowledges support from ANR Projet FREQUENCY (ANR-09-BLAN-0410-03). JA thanks the Royal Society for support in form of a Dorothy Hodgkin Research Fellowship. VV acknowledges financial support from the National Research Foundation and Ministry of Education in Singapore. VV is a fellow of Wolfson College Oxford.

\bibliographystyle{eptcs}
\bibliography{C:/DAMO/work/GenBib}

%\bibliographystyle{eptcs} % or whatever you prefer
%\begin{thebibliography}{1}
%
%\bibitem{GA08}
%R.J.~van Glabbeek, C.~Author \& Y.S.~Else (2008):
%\newblock \emph{An example of a paper with a rather large title-to-content ratio}.
%\newblock {\sl Electronic Proceedings in Theoretical Computer Science} 0,
% pp. 1--3.
%\newblock Available at \url{http://style.eptcs.org/}.

%\end{thebibliography}

\end{document}